\documentclass[a4paper,11pt]{article}
\usepackage{subcaption}
\usepackage{jheppub} % for details on the use of the package, please see the JINST-author-manual
\usepackage{lineno}
\usepackage{physics}
\usepackage{booktabs}
\usepackage{multirow}
\usepackage{xspace}

\newcommand{\DtoKspipi}{\ensuremath{D \rightarrow K^0_{\mathrm S}\pi^+\pi^-}\xspace} 
\newcommand{\DtoKsKK}{\ensuremath{D \rightarrow K^0_{\mathrm S} K^+ K^-}\xspace}
\newcommand{\xpm}{\ensuremath{x_\pm}\xspace}
\newcommand{\ypm}{\ensuremath{y_\pm}\xspace}
\newcommand{\xm}{\ensuremath{x_-}\xspace}
\newcommand{\ym}{\ensuremath{y_-}\xspace}
\newcommand{\xp}{\ensuremath{x_+}\xspace}
\newcommand{\yp}{\ensuremath{y_+}\xspace}

\newcommand{\rB}{\ensuremath{r_B}\xspace}
\newcommand{\dB}{\ensuremath{\delta_B}\xspace}

\newcommand{\ci}{\ensuremath{c_i}\xspace}
\newcommand{\si}{\ensuremath{s_i}\xspace}

\newcommand{\msqmin}{\ensuremath{m^2_-}\xspace}
\newcommand{\msqplus}{\ensuremath{m^2_+}\xspace}

\usepackage[T1]{fontenc}
\usepackage[scaled=0.85]{beramono}

\newcommand{\binscheme}[1]{\texttt{#1}\xspace}

\newcommand{\cleooptimal}{\binscheme{CLEO\_OPTIMAL}}

\newcommand{\newgamma}{\binscheme{NEWGAMMA}}
\newcommand{\newcharm}{\binscheme{NEWCHARM}}
\newcommand{\optkskkII}{\binscheme{OPT\_KSKK\_2}}
\newcommand{\optkskkIII}{\binscheme{OPT\_KSKK\_3}}
\newcommand{\optkskkIV}{\binscheme{OPT\_KSKK\_4}}
\newcommand{\equalVIII}{\binscheme{EQUAL\_8}}

\newcommand{\equalX}{\binscheme{EQUAL\_10}}

\usepackage{mathrsfs}

\title{\boldmath Optimal binning of $D\rightarrow K_{\mathrm S}^0\pi^+\pi^-$ and  $D\rightarrow K_{\mathrm S}^0K^+K^-$ phase space for experimental measurements}

\author[a]{M. Bovill}
\author[b]{N. Jurik}
\author[a]{S. Malde}
\affiliation[a]{University of Oxford,\\
Keble Road, United Kingdom}
\affiliation[b]{CERN, Geneva, Switzerland}

\emailAdd{marcelo.bovill@physics.ox.ac.uk}
\emailAdd{nathan.jurik@cern.ch}
\emailAdd{sneha.malde@physics.ox.ac.uk}

\abstract{New binning schemes for the Dalitz plots of \DtoKspipi and \DtoKsKK decays are presented. These are determined using a new figure of merit for optimisation that better represents the sensitivity to $\gamma$ than previous metrics. Further augmentation comes from including consideration of degeneracy with other physics parameters determined simultaneously and a more accurate treatment of relevant backgrounds. The overall expected improvement in the precision of $\gamma$ measurements using $B^\pm\to DK^\pm$ decays, followed by \DtoKspipi, is estimated at around 5$\%$. In addition, the first dedicated optimisation is performed for the observables relevant to charm mixing, $x_{CP}$ and $\Delta x$, in \DtoKspipi decays. This procedure accounts for both statistical sensitivity and the migration of events between bins due to detector resolution. The resulting binning scheme leads to an estimated gain of approximately 20$\%$ in statistical sensitivity, while maintaining the bias due to bin migration at a level comparable to the currently used scheme.}

\begin{document}
\maketitle
\flushbottom

\section{Introduction}

There are many measurements of key flavour observables that make use of the \DtoKspipi decay. These include determinations of the CKM angle $\gamma$~\cite{Run12_BPGGSZ, paper_Innes, paper_Seophine} and mixing in neutral charm mesons~\cite{Run2_binflip_prompt, Run2_binflip_sl,Belle:2024dux}. In these analyses, the \DtoKspipi Dalitz plot is studied in either a time-integrated or time-dependent manner to extract the physics observables of interest~\cite{GGSZ, Run12_BPGGSZ, Binflip}. The Dalitz plot encodes all kinematic information of the decay in two observables which are the squared invariant masses of the $K^0_{\mathrm{S}}\pi^+$ and $K^0_{\mathrm{S}}\pi^-$ pairs, denoted as $m_+^2$ and $m_-^2$, respectively. In order to extract the physics parameters of interest from these observed distributions, information on the $D$ decay strong-phase variation as a function of the position in the phase space of the $D$ decay, $(m_+^2,m_-^2)$, is required. On the Dalitz plot, the $CP$ conjugate final state of a point is given by the mirror point about the diagonal. Of particular interest is the strong-phase difference between $D^0$ and $\overline{D^0}$ decays to conjugate final kinematics, $\Delta\delta_D(m_+^2,m_-^2)=\delta_D(m_+^2,m_-^2)-\delta_D(m_-^2,m_+^2)$. While $\Delta\delta_D(m_+^2,m_-^2)$ can be described with an empirical amplitude model~\cite{BaBar2008, Belle2018, BaBar2010}, this introduces an undesirable model dependence, as the associated systematic uncertainties on the phase information are difficult to evaluate. The isobar formalism used by these amplitude models is known to violate unitarity~\cite{BaBar2008, Battaglieri:2014gca}, which is an unsolved problem in amplitude analyses to date, and from which a reliable strong-phase uncertainty cannot be propagated.

An alternative approach is to exploit the direct access to the strong-phase difference present in quantum-correlated $D^0\overline{D^0}$ pairs produced at threshold in $e^+e^-$ colliders. The measurement of strong-phase differences was pioneered by CLEO~\cite{old_optimal_binning}, with more precise measurements subsequently performed using the larger dataset available at BESIII~\cite{BESIII_8fb-1}. The Dalitz plot is partitioned into regions (bins), and the amplitude-weighted average cosine and sine of the strong-phase difference in each bin are measured. The schemes presented in Ref.~\cite{old_optimal_binning} have remained unchanged and the measurements of parameters related to the strong-phase difference have been used in a variety of different measurements~\cite{Run12_BPGGSZ, Run2_binflip_prompt, Run2_binflip_sl, paper_Innes, paper_Seophine, Run1_B0_GGSZ}.

The partitioning of the Dalitz plot into bins naturally dilutes the available statistical sensitivity, as variations in the strong-phase difference within the bin are ignored. In order to minimise this dilution, an optimisation metric that acts as a proxy for the statistical precision on $\gamma$ is defined using amplitude models. The bin boundaries are then determined such that this metric is maximised. It is stressed that the use of the model to determine the metric does not introduce model-dependence into the final measurement. The only result of differences between the models and a true description is that the statistical precision of the scheme is reduced with respect to the expectation. This leads to larger statistical uncertainties on $\gamma$ but introduces no bias on the central values or systematic uncertainty. 

This paper presents new optimisation metrics that improve the sensitivity to $\gamma$ when the \DtoKspipi decay is used. New binning schemes are also presented for the topologically similar \DtoKsKK decay. The same principles are then applied to improve sensitivity to charm mixing parameters. Since no dedicated optimisation for charm mixing parameters has been considered to date, the scope for improvement is larger. In addition to introducing the new metrics, it is timely to revisit these binning schemes for several reasons. A large 20$\,\mathrm{fb}^{-1}$ dataset has been collected by the BESIII experiment, but results for strong-phase parameters in \DtoKspipi and \DtoKsKK decays are yet to emerge. Providing new binning schemes now ensures they can be utilised in these forthcoming measurements. The larger dataset at BESIII motivates studying binning schemes with more bins, which can reduce the loss of statistical sensitivity. A large amount of experimental knowledge has been gained since the original binning schemes were devised. This includes a new amplitude model~\cite{Belle2018}, which agrees marginally better with the latest measurements from BESIII~\cite{BESIII_8fb-1}, as well as an improved understanding of experimental conditions and how the binning schemes can affect systematic uncertainties. This knowledge is used to inform choices on the new schemes with the aim of providing an overall better sensitivity to $\gamma$ and the charm mixing parameters.

Section~\ref{sec:methodology} describes the new metric used to optimise the binning scheme, the tools used to determine them, and the evaluation procedure for the potential improvements. Section~\ref{sec:gamma_binning} presents the choices made in determining a new scheme optimised for measuring $\gamma$ using the \DtoKspipi decay. Further changes to the optimisation metric that become necessary for the \DtoKsKK decay, and the resulting optimal schemes are subsequently presented in Sec.~\ref{sec:KsKK}. The ideas are then applied to measurements for charm mixing, and the resulting binning scheme is presented in Sec.~\ref{sec:charm_binning}.

\section{Methodology}
\label{sec:methodology}
The most powerful channel for measuring the CKM angle $\gamma$ is the $B^-\rightarrow DK^-$ decay and its charge conjugate process. Here, $D$ represents a superposition of $D^0$ and $\overline{D^0}$ mesons, which decay either to the $K_{\rm S}^0\pi^+\pi^-$ or $K_{\rm S}^0K^+K^-$ final states. The total $B$ decay amplitude is the sum of the favoured $B^- \to D^0K^-$ and suppressed $B^- \to \overline{D^0}K^-$ amplitudes. The magnitude of the ratio of these two amplitudes is given by $r_B$, while $\delta_B$ is the strong-phase difference between them. The probability density of the $B$ meson decays with the $D$ decay in a specific kinematic final state, $|A_{B^{-}}(m_{+}^2, m_{-}^2)|^2$, is given by
\begin{equation}
    \begin{split}
        |A_{B^{-}}(m_{+}^2, m_{-}^2)|^2 \propto
    &\Big(|A_{D}(m_{+}^2, m_{-}^2)|^2
        + r_B^2 |A_{D}(m_{-}^2,m_{+}^2)|^2 \\
            &+ 2|A_{D}(m_{+}^2, m_{-}^2)|
                |A_{D}(m_{-}^2, m_{+}^2)| \\
        &\quad\times
        \left[
        x_-\cos\Delta\delta_D(m_{+}^2, m_{-}^2)
        + y_-\sin\Delta\delta_D(m_{+}^2, m_{-}^2)
        \right]\Big),
    \end{split}
    \label{eq:A_B_signal}
\end{equation}
where $A_D(m_{+}^2, m_{-}^2)$ is the amplitude of the $D^0$ decay. Direct $CP$ violation in the $D$ decay is neglected so that the amplitude of the $\overline{D^0}$ decay is given by $A_D(m_{-}^2, m_{+}^2)$. Equation~\ref{eq:A_B_signal} gives sensitivity to $\gamma$ through the parameters $x_{-}$ and $y_{-}$, which are defined as
\begin{subequations}
    \begin{equation}
        x_{\pm}=r_B\cos(\delta_B\pm\gamma),
    \end{equation}
    \begin{equation}
        y_{\pm}=r_B\sin(\delta_B\pm\gamma).
    \end{equation}
\end{subequations}
The probability density of the $B^+$ decay can be found using Eq.~\ref{eq:A_B_signal} and making the substitutions $\msqplus \leftrightarrow \msqmin$, $\xm \rightarrow \xp$, and $\ym \rightarrow \yp$.
The $D$-decay phase space is partitioned into regions that are symmetric around $\msqmin=\msqplus$, resulting in $2 \times \mathcal{N}$ bins labelled from $i= -\mathcal{N}$ to $i= +\mathcal{N}$ (excluding zero). Bins for which $\msqmin > \msqplus$ are defined to have positive values of $i$. Integrating over the Dalitz-plot bin leads to the following parameters,
\begin{equation}
    K_i=\int_i |A_{D}(m_{+}^2, m_{-}^2)|^2\,\dd m_{+}^2\,\dd m_{-}^2,
    \label{eq:Ki}
\end{equation}
\begin{equation}
    c_i=\frac{1}{\sqrt{K_iK_{-i}}}\int_i |A_{D}(m_{+}^2, m_{-}^2)||A_{D}(m_{-}^2, m_{+}^2)|\cos\Delta\delta_D(m_{+}^2, m_{-}^2)\,\dd m_{+}^2\,\dd m_{-}^2,
    \label{eq:ci}
\end{equation}
where $K_i$ is proportional to the fraction of $D^0$ decays falling into bin $i$, and \ci is the amplitude-weighted cosine of the strong-phase difference. The quantity $s_i$ is defined analogously by replacing cosine with sine in Eq.~\ref{eq:ci}. Under the assumption that direct $CP$ violation in charm can be neglected, $c_i=c_{-i}$ and $s_i=-s_{-i}$.

Hence, integrating Eq.~\ref{eq:A_B_signal} over a Dalitz-plot bin gives the expected yield, $N_{i}^{B^{\pm}}$
\begin{equation}
    N_i^{B^{\pm}}
    = h_{B^{\pm}}\Big(K_{\mp i}
    + r_B^2 K_{\pm i}
    + 2\sqrt{K_i K_{-i}}
    \left(
    x_{\pm} c_i
    \mp y_{\pm} s_i
    \right)\Big),
    \label{eq:Ni_signal}
\end{equation}
where the $\pm$ and $\mp$ signs divide the equation into two forms according to the charge of the $B$ meson, and $h_{B^\pm}$ is a normalisation constant. To allow freedom in the positioning of bin boundaries, the Dalitz plot is discretised into an equally-spaced 500$\times$500 grid of sub-bins, ranging from 0.3 to 3.0 GeV$^2$. The sub-bins are assigned bin numbers indexed from 1 to $\mathcal{N}$ for $m_-^2>m_+^2$, while the bin-numbers for $m_-^2<m_+^2$ are defined anti-symmetrically. Sub-bins with centres that fall outside the kinematic limits of the decay are assigned a value of zero.

\subsection{Optimisation metric}
The bin numbers are assigned such as to maximise a chosen figure of merit, $Q$, that approximates the sensitivity to the observable of interest. Reference~\cite{old_optimal_binning} uses the form of the $Q$-metric first defined in Ref.~\cite{Bondar2008QuantumCorrelatedD0}, which built on the ideas of Ref.~\cite{Bondar2006}
\begin{equation}
    Q^2=\frac{\displaystyle{\text{\Large$\displaystyle\sum_{B^+,\,B^-}$}}{\text{\Large$\displaystyle\sum_{i=1}^\mathcal{N}$}}\frac{1}{N^{B^{\pm}}_i}\Bigg[\;\Bigg(\frac{\partial N^{B^{\pm}}_i}{\partial x_{\pm}}\Bigg)^2+\Bigg(\frac{\partial N^{B^{\pm}}_i}{\partial y_{\pm}}\Bigg)^2\;\Bigg]}{\displaystyle{\text{\Large$\displaystyle\sum_{B^+,\,B^-}$}}{\text{\Large$\displaystyle\int$}} \frac{1}{|A_{B^{\pm}}|^2}\Bigg[\;\Bigg(\frac{\partial|A_{B^{\pm}}|^2}{\partial x_{\pm}}\Bigg)^2+\Bigg(\frac{\partial|A_{B^{\pm}}|^2}{\partial y_{\pm}}\Bigg)^2\;\Bigg]\,\dd m_{+}^2\,\dd m_{-}^2}.
    \label{eq:Q_old_metric}
\end{equation}
The optimisation proceeds with the values $x_{\pm}=y_{\pm}=0$ so that Eq. \ref{eq:Q_old_metric} reduces to
\begin{equation}
    \displaystyle Q^2|_{r_B=0}=\frac{\displaystyle\sum_i K_i\bigg(c_i^2+s_i^2\bigg)}{\displaystyle\sum_i K_i}.
    \label{eq:Q_r_B_0}
\end{equation}

A binning scheme for \DtoKspipi decays determined using the metric in Eq.~\ref{eq:Q_r_B_0} and $\mathcal{N}=8$ was devised by CLEO and is referred to as the \cleooptimal scheme. It has been extensively used in measurements of $\gamma$~\cite{Run3_BPGGSZ,paper_Innes,paper_Seophine,Bellegamma, Belle_BelleII_BToDK}. The form of the metric defined in Eq. \ref{eq:Q_old_metric} is a measure of the statistical precision of $x_{\pm}$ and $y_{\pm}$ combined symmetrically in quadrature. However, the $x_{\pm}$ and $y_{\pm}$ parameters do not contribute equally to the sensitivity to $\gamma$, therefore this metric is only approximately representative of the true precision on $\gamma$. 

An alternative choice in the optimisation is to use values of \xpm and \ypm that are more representative than zero. This introduces a specific choice since there is a range of decay modes that are sensitive to $\gamma$, e.\;g. $B^0\to DK^{*0}$, $B^{\pm} \rightarrow D^{(*)} K^{(*)\pm}$. While $\gamma$ is common to all, they each have their own values of $\rB$ and $\dB$. The decay $B^{\pm}\rightarrow DK^{\pm}$ dominates the precision of the combination~\cite{2024_combination}. Therefore, tailoring the optimisation to this channel would overall be beneficial to the combination precision, as long as the sensitivity in other channels is not significantly reduced at the same time. In Ref.~\cite{DTo4Pi_binning}, the binning scheme for $D\rightarrow \pi^{\pm}\pi^{\mp}\pi^{\pm}\pi^{\mp}$ decays was tailored to $B^{\pm}\rightarrow DK^{\pm}$ by specifying the best-known values for $r_B$, $\delta_B$, and $\gamma$ in Eq.~\ref{eq:Q_old_metric}. To investigate this approach, signal-only pseudoexperiments are run on two binning schemes: one obtained with $x=y=0$, and the other with $x,y$ derived from the inputs in Ref.~\cite{2024_combination}. A negligible difference in the estimated statistical precision is observed between the two, which supports the idea that Eq. \ref{eq:Q_old_metric} does not accurately model the statistical precision on $\gamma$. 

In order to improve the sensitivity to $\gamma$, a new form of the metric is thus considered. This is based on the idea of using the Fisher information~\cite{Brandt} directly for $\gamma$. The Fisher information is a measure of the maximum amount of information that an observable provides on a variable of interest. A Poisson likelihood function of the form,
\begin{equation}
    \ln \mathcal{L}
    = \sum_{B^+,\,B^-}\sum_{i=-\mathcal{N}, i\neq 0}^{\mathcal{N}} \Bigl[ N_i^{{\rm obs}, B^{\pm}}\,\ln N_i^{B^{\pm}}\left(r_B,\delta_B,\gamma\right)
      - N_i^{B^{\pm}}\left(r_B,\delta_B,\gamma\right) - \ln\bigl(N_i^{{\rm obs}, B^{\pm}}!\bigr)\Bigr],
    \label{eq:log_likelihood}
\end{equation}
is considered, from which the Fisher information for $\gamma$ can be computed,
\begin{equation}
    \mathcal{I}(\gamma)
    \;=\; -\mathbb{E}\!\Bigg[\frac{\partial^2}{\partial\gamma^2}\ln\mathcal{L}\Bigg]
    \;=\; \sum_{B^+,\,B^-}\sum_{i=1}^{\mathcal{N}} \frac{1}{N_i^{B^{\pm}}}\Bigg(\frac{\partial N_i^{B^{\pm}}}{\partial\gamma}\Bigg)^2\geq\frac{1}{\mathrm{Var}(\gamma)},
    \label{eq:Fisher}
\end{equation}
where the inequality is the Cramér--Rao bound~\cite{Brandt}, which relates the Fisher information to the uncertainty on $\gamma$. The quantity defined in Eq. \ref{eq:Fisher} can then be used as the numerator for an improved $Q$-metric for statistical sensitivity to $\gamma$,
\begin{equation}
    Q^2_{\gamma}=
    \frac{\displaystyle
        \sum_{B^+,\,B^-}\sum_{i=1}^{\mathcal{N}}\,
        \frac{1}{N_i^{B^{\pm}}}\,
        \Bigg(\frac{\partial N_i^{B^{\pm}}}{\partial\gamma}\Bigg)^2
    }{\displaystyle
        \sum_{B^+,\,B^-}\int
        \frac{1}{|A_{B^{\pm}}|^2}\,
        \Bigg(\frac{\partial|A_{B^{\pm}}|^2}{\partial\gamma}\Bigg)^2
        \,\dd m_{+}^2\,\dd m_{-}^2
    },
    \label{eq:Q_new_metric}
\end{equation}
where the denominator is the Fisher information in the unbinned limit and
\begin{subequations}
    \begin{equation}
        \label{eq:derivatives}
        \begin{split}
            \frac{\partial \left|A_{B^{\pm}}\left(m_{+}^2, m_{-}^2\right)\right|^2}{\partial \gamma}
            \propto\;&
            \left|A_{D}\left(m_{+}^2, m_{-}^2\right)\right|
            \left|A_{D}\left(m_{-}^2, m_{+}^2\right)\right| \\
            &\times
            \left[
            - x_{\pm}\sin\Delta\delta_D\left(m_{+}^2, m_{-}^2\right)
            \mp y_{\pm}\cos\Delta\delta_D\left(m_{+}^2, m_{-}^2\right)
            \right],
        \end{split}
    \end{equation}
    \begin{equation}
        \frac{\partial N_i^{B^{\pm}}}{\partial\gamma}=2 h_{B^{\pm}}\sqrt{K_iK_{-i}}\left(-x_{\pm}s_i \mp y_{\pm}c_i\right).
        \label{eq:dNi_dgamma}
    \end{equation}
\end{subequations}
Hence, Eq. \ref{eq:Q_new_metric} represents the fraction of the unbinned Fisher information retained by a given binning scheme. Equivalently, via the Cramér--Rao bound, $Q_{\gamma}$ is the ratio of the minimum achievable statistical uncertainty on $\gamma$ in the unbinned case to that obtained with the chosen binning.

\subsection{Determining the optimised binning scheme}
The value of $Q_\gamma$ is determined for a given binning scheme from the amplitude model defined in Ref.~\cite{Belle2018}, where the values of amplitude and phase are taken at the sub-bin centre. Compared to the model used in Ref.~\cite{old_optimal_binning} the main differences are that the newer model has been developed on a dataset with a factor of 2.6 more $D$ meson decays and the addition of the Cabibbo-favoured and doubly-Cabibbo suppressed decays involving a $K^*(1410)$ resonance. A further advantage of this model is that it uses the value for the $D^0$ meson mass from the PDG2018~\cite{PDG2018}. This resolves a prevailing issue where $\simeq 0.3\%$ of $B$ meson data are excluded as the phase space of the currently available binning schemes is smaller than required.   

To reduce computation time, the optimisation routine is started from the equal-phase binning~\cite{old_optimal_binning, Bondar2006}, where the bins cover equal intervals of $\Delta\delta_D$. The bin numbers, $i$, in this case are defined as
\begin{equation}
    i(m_{+}^2, m_{-}^2)
    =
    1 + \left\lfloor
    \frac{\mathcal{N}}{2\pi}
    \left[
    \left(-\Delta\delta_D\left(m_{+}^2, m_{-}^2\right)
    + \frac{\pi}{\mathcal{N}}\right) \bmod 2\pi
    \right]
    \right\rfloor,
\end{equation}
for $m_+^2<m_-^2$. For $m_+^2>m_-^2$, the bin numbers $i$ are defined analogously but with opposite sign, $i(m_+^2,m_-^2)=-i(m_-^2,m_+^2)$. The optimisation algorithm proceeds via a random walk, where at each iteration the algorithm moves to the next sub-bin with a fixed step-size in a randomised direction. At each iteration, the considered sub-bin's bin number is reassigned, with the change kept if $Q$ increases, and discarded otherwise. In most iterations, the sub-bin is assigned the bin number of the sub-bin considered in the previous step. However, in a fraction of the iterations, $f_{\mathrm{grow}}$, the new bin number is assigned entirely at random, which allows for a new contiguous structure to ``grow'' within another. A value of 1$\%$ is chosen for $f_{\mathrm{grow}}$ along with a value of 3 sub-bins for the step size of the walk. This follows the algorithm used in Refs.~\cite{old_optimal_binning, Bondar2008QuantumCorrelatedD0}, with different choices for the random-walk parameters. The algorithm is run for a large number of iterations and terminates when further improvements in $Q_\gamma$ are less than 0.0001.

\subsection{Smoothing}

When optimising for a given metric, $Q$, small artefacts, such as an isolated sub-bin with a different bin number assignment to all its neighbours, can arise. To remove these, smoothing is applied to the binning schemes after optimisation. The procedure is based on that described in Ref.~\cite{old_optimal_binning}, in which the bin assignment of each sub-bin is compared with those within a local square region of radius, $r$, and is reassigned to the modal bin number if the fraction of the neighbouring sub-bins with the same bin assignment is less than a given threshold, $n_{\mathrm{cut}}$. The default values used here are $r=1$ and $n_{\textrm{cut}}$<0.23. The value of $r$ is much smaller than that in Ref.~\cite{old_optimal_binning}, since it is now known that the resolution in the Dalitz plot at LHCb is approximately the same size as a sub-bin~\cite{Mikkel_thesis, Hilton:2021thh}. Nominally, this procedure is run iteratively until no further changes. The typical loss in the value of the $Q$ metric before and after smoothing is 0.003 or less. 

\subsection{Pseudoexperiments}
\label{sec:toys}
Pseudoexperiments are run to evaluate the binning schemes in a consistent and accurate fashion amongst the various forms of the $Q$-metric considered in the optimisation.
The expected yields for these pseudodatasets, $N_i^{B^{\pm}}$, are generated using Eq. \ref{eq:Ni_signal}, where the $K_i$, $c_i$, and $s_i$ are computed from the amplitude model~\cite{Belle2018} and the values of $r_B$, $\delta_B$, and $\gamma$ are taken from the latest LHCb combination~\cite{2024_combination}. It is assumed that $h_{B^{+}}=h_{B^{-}}$. Statistical fluctuations are then added to the $N_i^{B^{\pm}}$, sampled according to a Poisson distribution. 

These pseudodatasets are analysed using a two-step process. In the first step, the $x_{\pm}$ and $y_{\pm}$ parameters are determined by a fit to the $N_i^{B^{\pm}}$ distribution using a Poisson likelihood. In the second stage, the values of $x_{\pm}$ and $y_{\pm}$ and their covariance determine the physics parameters, $r_B$, $\delta_B$, and $\gamma$, via a $\chi^2$ fit. The width of the resulting distribution of fitted $\gamma$ values is then taken as a measure of the statistical uncertainty on $\gamma$. In order to resolve the differences in sensitivity between binning schemes, the number of events generated in a given pseudoexperiment and the number of pseudoexperiments run are sufficiently large such that the statistical uncertainty in the width of the fitted parameter values is much smaller than the differences between the binning schemes.

\section{\texorpdfstring{Optimal binning for sensitivity to $\gamma$}{Optimal binning for sensitivity to gamma}}
\label{sec:gamma_binning}

\subsection{Signal-only optimisation}
\label{sec:Q_metric}

Pseudoexperiments with only signal events are used to verify the improvement in sensitivity when optimising using Eq. \ref{eq:Q_new_metric} rather than Eq. \ref{eq:Q_r_B_0}. The results are presented normalised to the \cleooptimal scheme in Table~\ref{tab:toy_results_summary}. Hence, values less than one describe an improvement in sensitivity. An improvement of 5.7$\%$ relative to the \cleooptimal binning scheme is seen. Table~\ref{tab:toy_results_summary} also shows the improvement in sensitivity as $\mathcal{N}$ is increased. Successive gains of between 1$\%$ and 2$\%$ are observed when $\mathcal{N}$ is increased from 8 to 10 and from 10 to 12.
\begin{table}[ht!]
    \centering
    \caption{Relative sensitivities to $\gamma$ in pseudoexperiments optimised using Eq.~\ref{eq:Q_new_metric} with different values of $\mathcal{N}$ with respect to the \cleooptimal scheme. Also shown are the corresponding values of $Q_{\gamma}$. The statistical uncertainties on the ratio are $\sim 0.003$.}
    \label{tab:toy_results_summary}
    \begin{tabular}{ccc}
       \toprule
       $\mathcal{N}$ & $\sigma_{\gamma}/\sigma_{\gamma}^{\cleooptimal}$ & $Q_{\gamma}$ \\
       \midrule
       8  & 0.943 & 0.912 \\
       10 & 0.932 & 0.926 \\
       12 & 0.915 & 0.941 \\
       15 & 0.899 & 0.953 \\
       \bottomrule
    \end{tabular}
\end{table}
While it is expected that the sensitivity should improve as $\mathcal{N}$ is increased, the choice of how many bins to use is dependent on the datasets that will be used to make the measurements of \ci and \si, as well as the $B$ meson data used to make the $CP$ violation measurement. The dataset available at BESIII is $\sim 25$ times larger than that collected at CLEO~\cite{BESIII_integrated_luminosity, old_optimal_binning}. However, the measurement itself has become more complex with the removal of the weak-model constraint~\cite{BESIII_8fb-1}. Furthermore, the \DtoKspipi decay is not just a signal channel in strong-phase measurements; it also functions as a valuable tag for strong-phase measurements in other decays which may also be binned more finely than they have been to date, e.\;g. $D \to K^\mp\pi^\pm\pi^{\mp}\pi^\pm$~\cite{BESIII_K3Pi:2026}. Hence, the measurement at BESIII will become problematic for large $\mathcal{N}$. On the $B$ meson side, measurements of $\gamma$ have been shown to suffer small biases due to the partitioning of data leading to low per-bin yields~\cite{Run12_BPGGSZ, paper_Innes, paper_Seophine}. However, the size of these datasets is expected to increase in the coming decade. Therefore, balancing the gain in statistical precision with these concerns, the scheme with $\mathcal{N}=10$ is chosen as the nominal.

\subsection{Backgrounds}
\label{sec:backgrounds}

Reference~\cite{old_optimal_binning} showed that the presence of background reduces the expected precision achievable with the binning scheme, as the high levels of asymmetry in small bins are washed out by the background. Therefore, the best binning scheme for a given decay mode will also take into account the background level and distribution over the Dalitz plot. This was investigated in Ref.~\cite{old_optimal_binning} where a dedicated scheme that took background effects into account was presented. However, the levels of background used were far higher than those subsequently observed at LHCb or Belle II, which rendered this scheme of limited use. The inclusion of background is revisited in light of better knowledge of experimental conditions.

Two background types with distinct distributions on the Dalitz plot are considered. The first is composed of real $D^0$/$\overline{D^0}$ decays paired with a random kaon. This background type is assumed to be composed of equal proportions of $D^0$ and $\overline{D^0}$ mesons. The second background type is purely combinatorial and is assumed to be distributed evenly over the Dalitz plot. Both background PDFs are assumed to be $CP$-invariant, while the signal distribution violates $CP$ symmetry. The signal and background PDFs are normalised such that
\begin{subequations}
\label{eq:PDF_normalisation}
    \begin{equation}
        \displaystyle\sum_{B^+,\,B^-}\int\Big|A_{B^{\pm}}(m_{+}^2, m_{-}^2)\Big|^2\,\dd m_{+}^2\,\dd m_{-}^2=\sum_{B^+,B^-}\sum_i N_i^{B^{\pm}}=1,
    \end{equation}
    \begin{equation}
        \displaystyle\int \mathcal{B}^c(m_{+}^2, m_{-}^2)\,\dd m_{+}^2\,\dd m_{-}^2=\sum_i \mathcal{B}_{i}^c=1,
    \end{equation}
\end{subequations}
where $c$ is the background type and $\mathcal{B}_{i}^c$ is the $c$th background PDF integrated over bin $i$. The distribution of observed events on the Dalitz plot and observed bin yields become
\begin{subequations}
\label{eq:Ni_modified}
    \begin{equation}
        \Big|A^{\prime}_{B^{\pm}}(m_{+}^2, m_{-}^2)\Big|^2=f\Big|A_{B^{\pm}}(m_{+}^2, m_{-}^2)\Big|^2+\frac{1}{2}\sum_c \tilde{f}^c\mathcal{B}^c(m_{+}^2, m_{-}^2),
        \label{eq:A_B_modified}
    \end{equation}
    \begin{equation}
        N_i^{\prime B^{\pm}}=f N_i^{B^{\pm}}+\frac{1}{2}\sum_c \tilde{f}^c\mathcal{B}_i^c,
        \label{eq:Ni_binned_modified}
    \end{equation}
\end{subequations}
respectively, where $f$ and $\tilde{f}^c$ are the fractions of the charge-integrated yields of the signal and $c$th background component which obey the relation, $f+\sum_c \tilde{f}^c=1$. The total background is evenly split across the charges, hence the factor of $\frac{1}{2}$ in Eq.~\ref{eq:Ni_modified}. In both the binning scheme optimisation and generation of the pseudodatasets, the $N_i^{B^{\pm}}$ in Eqs.~\ref{eq:log_likelihood} and \ref{eq:Q_new_metric} are replaced by $N_i^{\prime B^{\pm}}$ from Eq.~\ref{eq:Ni_binned_modified}. In this way, the different bins are effectively re-weighted in the optimisation metric according to their signal purity.

The choice is made to optimise based on the background levels present at the LHCb experiment. Data at LHCb are divided into two categories, \emph{Long} and \emph{Downstream}, based on where the $K_{\mathrm{S}}^0$ meson decays in the detector. Each category has a different background distribution. The relative amount of background to signal is taken from the fitted yields reported in LHCb Runs 1 and 2~\cite{Run12_BPGGSZ} and the division of the background into the two categories is determined from Ref.~\cite{Mikkel_thesis}. These background levels are shown in Table~\ref{tab:background_levels}. The nominal background levels used in the optimisation and pseudoexperiment generation are the weighted average of the two categories and are shown in bold in Table~\ref{tab:background_levels}.

\begin{figure}[ht!]
    \centering
    \begin{subfigure}[b]{0.49\textwidth}
        \centering
        \includegraphics[width=\textwidth]{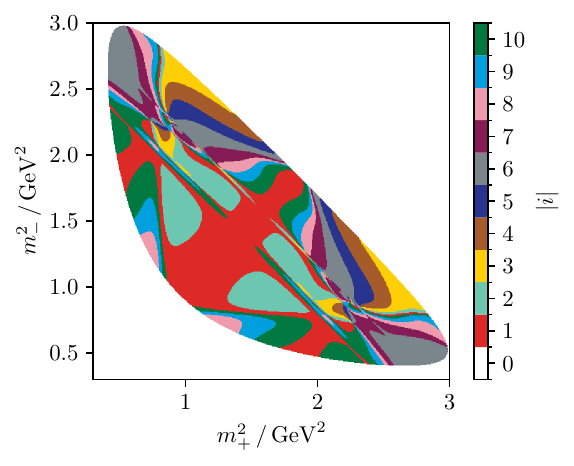}
        \caption{$\mathcal{N}=10$ equal-phase scheme, $Q_{\gamma}=0.833$}
    \end{subfigure}
    \hfill
    \begin{subfigure}[b]{0.49\textwidth}
        \centering
        \includegraphics[width=\textwidth]{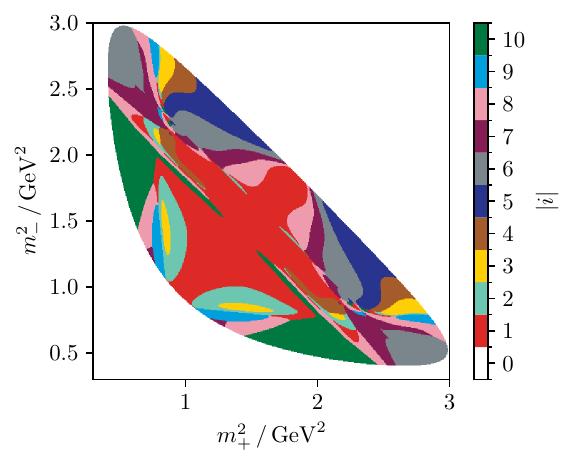}
        \caption{$\mathcal{N}=10$ \newgamma scheme, $Q_{\gamma}=0.942$}
    \end{subfigure}
    \caption{The \newgamma binning scheme optimised for statistical precision on $\gamma$, with backgrounds present, alongside an equal-phase scheme for comparison.}
    \label{fig:gamma_binning}
\end{figure}

\begin{table}[ht!]
    \centering
    \caption{The values of $\tilde{f}^c$ determined from information in Refs.~\cite{Run12_BPGGSZ,Mikkel_thesis}. The weighted average, shown in bold, is taken as the nominal.}
    \label{tab:background_levels}
    \small
    \begin{tabular}{cccc}
      \toprule
        $\mathcal{B}^c$ & \emph{Long} & \emph{Downstream} & Nominal \\
      \midrule
        Real-$D$ & 0.044 & 0.026 & \textbf{0.032}\\
        Comb. & 0.007 & 0.023 & \textbf{0.018} \\
        \bottomrule
   \end{tabular}
\end{table}
Similarly to Sec.~\ref{sec:Q_metric}, the effect of backgrounds on the sensitivity to $\gamma$ is evaluated with pseudoexperiments. 
Relative sensitivities using schemes optimised with and without the addition of background are shown in Table~\ref{tab:toy_results_backgrounds}. Both schemes use $\mathcal{N}=10$ and the results are presented normalised to the sensitivity using the \cleooptimal scheme. The pseudodatasets are generated with background. The results indicate that the improvements of $7\%$ shown in Table~\ref{tab:toy_results_summary} are not achieved once background is factored in. If the optimisation assumes no background then the gain is 3.8$\%$. This gain recovers to 5$\%$ when the optimisation takes into account the background distribution, demonstrating the value of incorporating background effects. The $Q_{\gamma}$ value implies that the \newgamma scheme retains 94$\%$ of the statistical sensitivity compared to the unbinned limit for \DtoKspipi decays, which is verified in pseudoexperiments where the unbinned limit is approximated by treating each sub-bin as its own bin. The scheme including background is put forward as the optimal scheme for measuring $\gamma$ and is named \newgamma. The scheme is visualised in Fig.~\ref{fig:gamma_binning}, alongside the equal-phase binning with $\mathcal{N}=10$ (\equalX).

\begin{table}[ht!]
    \centering
    \caption{Statistical sensitivities for $\gamma$, using different optimisation conditions, relative to the \cleooptimal scheme. The statistical uncertainties on the ratio are $\sim 0.003$.}
    \label{tab:toy_results_backgrounds}
    \begin{tabular}{cll}
    \\
    
       \toprule
      Scheme Name & Optimisation conditions & $\sigma_\gamma/\sigma_{\gamma}^\cleooptimal$ \\
-& $\mathcal{N}=10,\ \tilde{f}^c=0$ & 0.962 \\
\newgamma & $\mathcal{N}=10,\ \tilde{f}^c=$nominal & 0.950 \\
       \bottomrule
    \end{tabular}
\end{table}

To check that the scheme is robust at a range of realistic background levels, different from those used in the optimisation, pseudoexperiments are generated with a variety of different background levels and distributions. These include the levels observed in LHCb Run 3~\cite{Run3_BPGGSZ}. In all tested scenarios, the sensitivity is better using the \newgamma scheme than the scheme optimised for the signal-only case with $\mathcal{N}=10$. In the slightly higher background levels of Run 3, the relative gain in using \newgamma is found to be slightly larger than in the nominal case. This confirms that the \newgamma scheme will remain useful with realistic variations to the background levels that may occur in the future.

\subsection{Detector efficiencies}
\label{sec:detector_efficiencies}

Another effect that can have an impact is the aggregate detector efficiency, which is expected to be non-uniform across the Dalitz plot. The effects of detector efficiency are incorporated into Eqs.~\ref{eq:A_B_modified} and~\ref{eq:Ni_binned_modified} by replacing the amplitudes, $|A_D(m_+^2,m_-^2)|$, with
\begin{equation}
    A_D^{\prime}(m_+^2,m_-^2)=A_D(m_+^2,m_-^2)\,\sqrt{\epsilon(m_+^2,m_-^2)},
    \label{eq:A_D_efficiencies}
\end{equation}
where $\epsilon(m_+^2,m_-^2)$ is the relative detector efficiency from one point on the Dalitz plot to another. The efficiency is treated as symmetric about the diagonal of the Dalitz plot. The model for the efficiency variation is taken from Ref.~\cite{cheung_2017_t51et-en633}, where the efficiencies from simulation of \DtoKspipi at LHCb, based on a variety of different selection criteria, are available. Binning schemes incorporating the different efficiency profiles into the metric are determined, and the resulting sensitivity evaluated using pseudoexperiments. The resulting improvement in sensitivity is found to be marginal. Furthermore, the efficiency profile is highly dependent on the data selection criteria, which are subject to change in the future. Hence, given the negligible gain and the uncertainty in the future efficiency profile, this addition to the optimisation metric is not considered further.  

\subsection{Implications for systematic uncertainties when used in analysis}
\label{sec:systematics}
\label{sec:strong_phase_systematic}
It is important to verify that the \newgamma scheme does not have unintended effects on the systematic uncertainties of the analysis. For instance, the uncertainties on $c_i$ and $s_i$ propagated to the measurement of $\gamma$~\cite{Run12_BPGGSZ} already constitute the largest systematic uncertainty in recent analyses~\cite{Run12_BPGGSZ, Run3_BPGGSZ, paper_Innes, paper_Seophine}, and this contribution may eventually become a limiting factor as datasets grow. The impact on this uncertainty due to the proposed change in binning scheme is therefore assessed. The propagated uncertainty on $\gamma$ is dominated by the contribution from the uncertainty on \si, since they are known less precisely than the \ci. The value of \si is primarily determined from data at BESIII where both $D$ mesons decay to the \DtoKspipi final state~\cite{BESIII_8fb-1}. The yield of events where one \DtoKspipi decays to bin $i$ and the other decays to bin $j$, $N_{ij}$, is given by
\begin{equation}
    N_{ij}=\displaystyle h\left[K_{i}K_{-j}+K_{-i}K_{j}-2\sqrt{K_iK_{-i}K_jK_{-j}}\left(c_ic_j+s_is_j\right)\right],
    \label{eq:double_tag_yields}
\end{equation}
where $h$ is a normalisation factor. A proxy for the statistical uncertainty on \si is determined in the following way. Pseudodatasets are generated according to Eq.~\ref{eq:double_tag_yields} where the bin yields are sampled according to a Poisson distribution. In the pseudodataset generation, $h$ is set such that $\sum_{ij} N_{ij}$ is the total number of expected signal events obtained by linearly scaling the reported yields in the 8 fb$^{-1}$ dataset~\cite{BESIII_8fb-1} to 20 fb$^{-1}$ which is the size of the dataset currently available at BESIII~\cite{BESIII_integrated_luminosity}.

The generated pseudodata are fit to determine the \si values. 
The fit imposes the physical constraint $c_i^2+s_i^2<1$, and since the measurements of $c_i$ are well constrained from other decay channels at BESIII, a tight Gaussian constraint is applied to the $c_i$ such that they are effectively fixed. The fitted values are then used to determine a propagated uncertainty on $\gamma$ in a similar fashion to that described in Ref.~\cite{Run12_BPGGSZ}.
The same study is run for the \cleooptimal binning scheme such that a relative comparison can be made. It is found that the expected propagated systematic uncertainty is approximately $10\%$ smaller in the \newgamma binning scheme. This is an initially surprising result since the increased number of bins should result in larger statistical uncertainties on \si. The same study is run using binning schemes optimised for a variety of $\mathcal{N}$, and the results are presented in Table~\ref{tab:strong_phase_systematic}. The propagated uncertainty is found to decrease as $\mathcal{N}$ is increased. Further study reveals that, as $\mathcal{N}$ increases, the expected values of \ci and \si move closer to the unit circle. It is this change that reduces the systematic uncertainty. A similar effect is observed when comparing the propagated uncertainties in Refs.~\cite{Run12_BPGGSZ} and~\cite{Run3_BPGGSZ}, where the uncertainties in the latter increase due to the difference in the central values of the strong-phase parameters. Therefore, it is concluded that the new binning scheme will not have an adverse impact on the propagated uncertainty, and under the assumption that the amplitude model is correct, it could deliver a $10\%$ improvement.

\begin{table}[ht!]
   \centering
   \caption{The ratio of propagated uncertainties due to strong-phase inputs between schemes optimised with Eq. \ref{eq:Q_new_metric} and \cleooptimal.}
   \label{tab:strong_phase_systematic}
   \begin{tabular}{ccccc}
      \toprule 
      & $\mathcal{N}=5$ & $\mathcal{N}=8$ & $\mathcal{N}=10$ & $\mathcal{N}=12$ \\
&      & &\newgamma & \\
\midrule

$\sigma_{\rm syst}/\sigma_{\rm syst}^\cleooptimal$ & 0.927 & 0.918 & 0.906 & 0.892 \\
      \bottomrule
   \end{tabular}
\end{table}

Another systematic uncertainty on $\gamma$ that is related to the specifics of the binning scheme arises from bin migration where events are reconstructed in different bins from their true bin due to the smearing of the particle momentum. The measurement of $\gamma$ considers only the observed bin yields $N_i^{B^{\pm}}$, and therefore it is only the net migration relative to the bin population, $M_i$, into each bin that is relevant to this uncertainty.

\begin{figure}
    \centering
    \includegraphics[width=\linewidth]{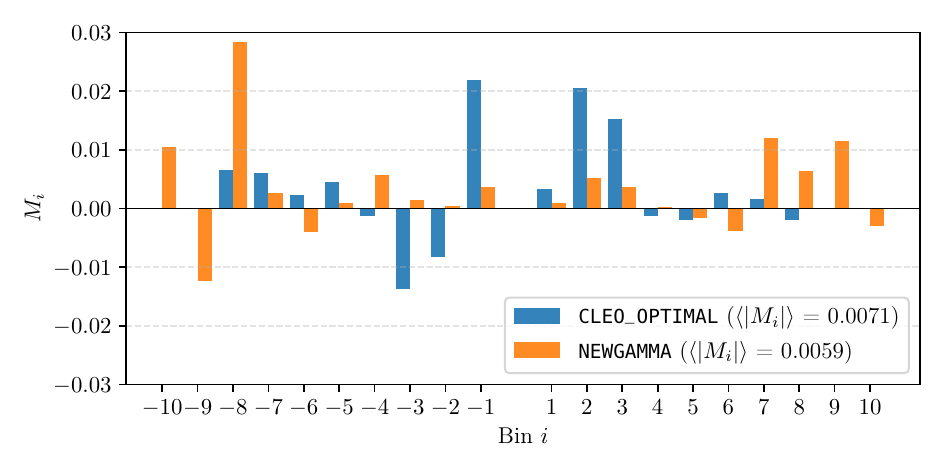}
    \caption{Estimated net migrations for the \cleooptimal and \newgamma schemes.}
    \label{fig:net_migration_gamma}
\end{figure}

The detector resolution is modelled by a Gaussian, with a width of 0.0085 GeV$^2/c^4$ in $m^2_{\pm}$ and a correlation coefficient of -0.6 between them. These parameters are chosen conservatively based on the resolutions reported in Refs.~\cite{Hilton:2021thh, Mikkel_thesis}. To estimate the $M_i$, a large number of events are generated according to the amplitude model~\cite{Belle2018} to assign the true position, with the reconstructed positions computed by adding shifts sampled from the resolution function. The resulting net migrations, $M_i$, are displayed in Fig.~\ref{fig:net_migration_gamma} for the \cleooptimal and \newgamma schemes, from which it can be seen that the average magnitude of $M_i$ is smaller in the \newgamma scheme. As a further test, a more detailed study similar to that described in Ref.~\cite{Run3_BPGGSZ} is carried out, which confirms that the systematic effect due to migration remains small in the \newgamma scheme.

\subsection{\texorpdfstring{Optimal binning of \DtoKsKK for sensitivity to $\gamma$}{Optimal binning of D to KsKK for sensitivity to gamma}}
\label{sec:KsKK}

Another $D$ decay that is used for $\gamma$ measurements is \DtoKsKK. While it has a smaller branching fraction than \DtoKspipi decays, it is often analysed simultaneously due to the similarities in their topology~\cite{paper_Innes, paper_Seophine, Run12_BPGGSZ} and improves the precision on $\gamma$ by approximately 10$\%$ compared to using only \DtoKspipi decays. Reference~\cite{old_optimal_binning} presents three equal-phase binning schemes for \DtoKsKK decays, corresponding to $\mathcal{N}=2, 3$, and $4$. As in Ref.~\cite{old_optimal_binning}, the amplitude model of Ref.~\cite{BaBar2010} is adopted in this study. Only the two-bin variant has been used to date for $\gamma$ measurements, although the accumulated dataset at LHCb could allow a finer binning scheme to be used.

An optimisation using Eq.~\ref{eq:Q_new_metric} is performed and the results are validated with pseudoexperiments. It is observed that, unlike for the \DtoKspipi channel, there are large discrepancies between the relative changes in $Q_{\gamma}$ and the expected sensitivities in pseudoexperiments for the schemes with $\mathcal{N}=2, 3$, and $4$. This suggests that Eq.~\ref{eq:Q_new_metric} does not describe the sensitivity to $\gamma$ well in these schemes. This coincides with a much stronger correlation between the determined values of $\gamma$ and $\dB$ than in the \DtoKspipi schemes, particularly in the schemes with $\mathcal{N}=2$. This correlation arises due to the smaller $\mathcal{N}$ used to determine $\gamma$. The new metric defined in Eq.~\ref{eq:Q_new_metric} uses the univariate Fisher information and, therefore, ignores this correlation. This motivates the use of a metric that better represents the sensitivity to $\gamma$ by taking these correlations into account. This is achieved using a Fisher information matrix defined as
\begin{equation}
  [\mathcal{I}] =
  \begin{pmatrix}
    \mathcal{I}_{\gamma\gamma}    & \mathcal{I}_{\gamma r_B}    & \mathcal{I}_{\gamma\delta_B}    \\
    \mathcal{I}_{r_B\gamma}       & \mathcal{I}_{r_B r_B}       & \mathcal{I}_{r_B\delta_B}       \\
    \mathcal{I}_{\delta_B\gamma}  & \mathcal{I}_{\delta_B r_B}  & \mathcal{I}_{\delta_B\delta_B}
  \end{pmatrix},
  \label{eq:FisherMatrix}
\end{equation}
where
\begin{equation}
    \mathcal{I}_{jk}
    =
    -\mathbb{E}\bigg[
    \frac{\partial^2 \ln \mathcal{L}}
    {\partial\lambda_j \partial\lambda_k}
    \bigg],
    \label{eq:Ijk}
\end{equation}
and $\lambda$ is the vector of parameters. Then $\mathcal{I}_{\gamma\gamma}=\mathcal{I}(\gamma)$ is the univariate term defined in Eq.~\ref{eq:Fisher}. The effective Fisher information for $\gamma$, marginalised over $(r_B,\delta_B)$, follows from the Cramér--Rao bound and is given by the reciprocal of the (1,1) element of Eq.~\ref{eq:FisherMatrix},
\begin{equation}
  \mathcal{I}_{\text{eff},\gamma}
  = \bigl[(\mathcal{I}^{-1})_{\gamma\gamma}\bigr]^{-1}.
  \label{eq:Ieff}
\end{equation}
Analogously to Eq. \ref{eq:Q_new_metric}, a $Q$-metric can be defined,
\begin{equation}
    Q^2_{\textrm{eff},\gamma}=\frac{\mathcal{I}_{\text{eff},\gamma}}{\mathcal{I}^{\textrm{unbinned}}_{\text{eff},\gamma}}.
    \label{eq:Q_eff}
\end{equation}
It can be seen from Eq.~\ref{eq:Ijk} that, since $\frac{\partial N_i^{B^\pm}}{\partial\delta_B}=\pm\frac{\partial N_i^{B^\pm}} {\partial\gamma}$, the off-diagonal elements $\mathcal{I}_{\gamma\delta_B}$ and $\mathcal{I}_{\delta_{B}\gamma}$ are large. This explains the strong correlations between the fitted $\gamma$ and $\delta_B$ values observed in pseudoexperiments for schemes with low $\mathcal{N}$. Hence, the additional terms in Eq. \ref{eq:Q_eff} involving the off-diagonal terms of Eq. \ref{eq:FisherMatrix} change the optimisation metric. 

In Table~\ref{tab:d2kskk}, results of sensitivity to $\gamma$ are presented for the $\mathcal{N}=2, 3$, and $4$ cases where the binning scheme has been determined by the equal-phase binning, the metric given in Eq.~\ref{eq:Q_new_metric} and the metric given in Eq.~\ref{eq:Q_eff}. For the latter optimal schemes, backgrounds are part of the optimisation using the yields and composition reported in Ref.~\cite{Run12_BPGGSZ,Mikkel_thesis}, analogously to the studies for \DtoKspipi decays. The most dominant contribution to the sensitivity is the value of $\mathcal{N}$ used, where a 10$\%$ gain comes from moving from $\mathcal{N}=2$ to 3. The use of Eq.~\ref{eq:Q_new_metric} provides a small gain compared to the equal-phase binning scheme with the same value of $\mathcal{N}$. The metric in Eq.~\ref{eq:Q_eff} makes further improvement, in particular for $\mathcal{N}=2$. It is verified that the form of the metric in Eq.~\ref{eq:Q_eff} does not provide significant improvement in the sensitivity to $\gamma$ for \DtoKspipi decays. As expected, the larger $\mathcal{N}$ results in a minimal difference between the metrics in Eqs.~\ref{eq:Q_new_metric} and \ref{eq:Q_eff}.

\begin{table}[ht!]
    \caption{Relative sensitivities for the $\mathcal{N}=2$, 3, and 4 equal-phase binning schemes and schemes optimised for Eq. \ref{eq:Q_new_metric} and Eq. \ref{eq:Q_eff}. Backgrounds are present in both the optimisation and generation of pseudoexperiments. The sensitivities are normalised to the two-bin equal-phase scheme.}
    \label{tab:d2kskk}
    \centering
    \begin{tabular}{cccc}
    \\
    \toprule
         Optimisation metric& Equal & Eq.~\ref{eq:Q_new_metric} & Eq.~\ref{eq:Q_eff} \\ \midrule
        $\mathcal{N}=2$ & 1 & 0.988 & 0.969 \\
        $\mathcal{N}=3$ & 0.906 & 0.876 & 0.872 \\
        $\mathcal{N}=4$ & 0.885 & 0.818 & 0.808 \\ \bottomrule
    \end{tabular}
\end{table}

\begin{figure}[ht!]
    \centering
    \begin{subfigure}[b]{0.32\textwidth}
        \centering
        \includegraphics[width=\textwidth]{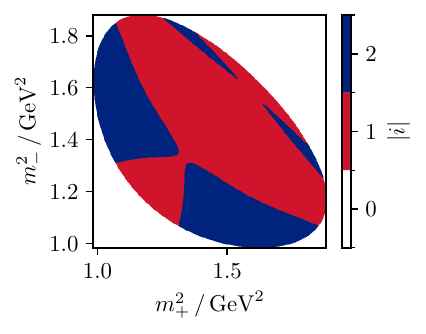}
        \caption{\optkskkII, $Q_{\textrm{eff},\gamma}=0.687$}
    \end{subfigure}
    \begin{subfigure}[b]{0.32\textwidth}
        \centering
        \includegraphics[width=\textwidth]{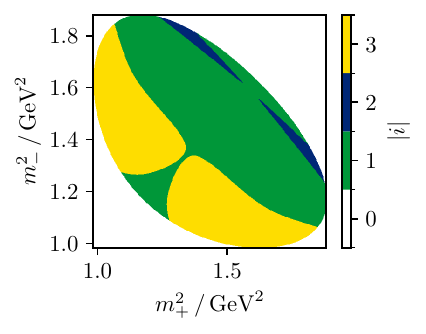}
        \caption{\optkskkIII, $Q_{\textrm{eff},\gamma}=0.785$}
    \end{subfigure}
    \begin{subfigure}[b]{0.32\textwidth}
        \centering
        \includegraphics[width=\textwidth]{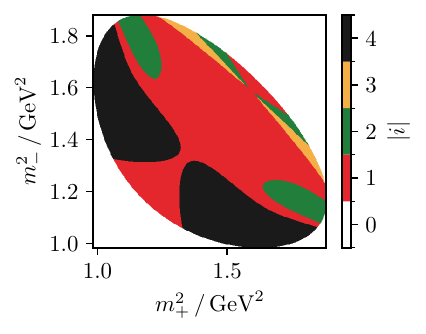}
        \caption{\optkskkIV, $Q_{\textrm{eff},\gamma}=0.853$}
    \end{subfigure}
    \caption{The \optkskkII, \optkskkIII, and \optkskkIV binning schemes for the \DtoKsKK channel, along with their values of $Q_{\textrm{eff},\gamma}$.}
    \label{fig:kskk_binning}
\end{figure}

The three schemes for the \DtoKsKK channel from the metric in Eq.~\ref{eq:Q_eff} are presented in Fig.~\ref{fig:kskk_binning}, and are labelled the \optkskkII, \optkskkIII, and \optkskkIV schemes. It is noted that, as in the previous equal-phase $\mathcal{N}=3$ and $4$ schemes~\cite{old_optimal_binning}, the \optkskkIII and \optkskkIV contain bins of relatively low $K_i$. Coupled with the low branching fraction of the \DtoKsKK decay~\cite{PDG2014}, schemes of larger $\mathcal{N}$ are therefore not considered, since measuring the $c_i$ and $s_i$ with reasonable precision would likely not be feasible.

Similarly to the \DtoKspipi channel, the systematic uncertainty due to the measurement uncertainties of $c_i$ and $s_i$ is studied in pseudoexperiments. It is found to be slightly reduced in the \optkskkIII and \optkskkIV schemes with respect to the equal-phase $\mathcal{N}=2$ scheme. The systematic uncertainty due to bin migration is also studied, and found to be slightly larger in the \optkskkIII scheme than in the equal-phase $\mathcal{N}=2$ scheme. However, since the size of the systematic uncertainty is sub-dominant compared to other systematic uncertainties and the \DtoKsKK decay has a sub-leading contribution to sensitivity to $\gamma$~\cite{Run12_BPGGSZ, Run3_BPGGSZ}, this increase is acceptable in light of the possible statistical gains.

\subsection{Implications of the proposed binning schemes}
A binning scheme that is optimal for $\gamma$ measurements in $B^\pm\to D K^\pm$ decays, referred to as the \newgamma scheme, is presented in Fig.~\ref{fig:gamma_binning}. Choices have been made regarding the specific $B$ decay and the expected background levels at the LHCb experiment. Schemes are not presented for a variety of $B$ decays or typical background levels at different experiments. The ultimate sensitivity to $\gamma$ comes from a combination of many measurements. This combination should take into account the repeated use of the strong-phase measurements, which is only possible if all measurements using the \DtoKspipi channel use the same binning scheme. The proposal in this paper is therefore that the \newgamma scheme should be used for all $B$ decays. The impact on $B^0 \to D K^{\star 0}$ decays is assessed in pseudoexperiments. This channel has a value of $r_B$ that is three times larger than for $B^\pm \to DK^\pm$ and higher levels of combinatorial background~\cite{paper_Innes}, and hence has properties different from the optimisation choices. When tested on the \newgamma scheme, it is found that the precision is no worse than that of the \cleooptimal scheme, which validates the choice to maximise the sensitivity to the $B^\pm \to D K^\pm$ decay.

The CKM angle $\gamma$ has also been measured at the Belle and Belle II experiments~\cite{Run12_BPGGSZ} using the \cleooptimal scheme. While the background levels are higher than at LHCb, its composition is dominated by background originating from the misidentification of pions as kaons, which has less impact on the sensitivity to $\gamma$. Given the finding in Sec.~\ref{sec:backgrounds} that the gain in sensitivity was robust across a range of background scenarios, it is expected that the \newgamma scheme will maintain a gain in statistical precision over the \cleooptimal scheme under the experimental conditions in Belle/Belle II, although the gain may not be as large as in the nominal case in LHCb conditions considered here. 

In the \DtoKsKK channel, three schemes are presented. However, for control over systematic uncertainties in the combination of a variety of different $\gamma$ measurements, it is essential that one common scheme is used for the \DtoKsKK decay. The relatively small per-bin yields $K_i$ in the $\mathcal{N}=4$ scheme may make precise measurements of \si challenging. However, the \optkskkIII scheme is viewed as a good compromise in terms of substantially improved statistical precision compared to the current $\mathcal{N}=2$ scheme while being expected to present fewer complications for measurements of the strong-phase parameters than the \optkskkIV scheme.

Several unbinned model-independent methods for measuring $\gamma$ have been proposed~\cite{Poluektov:2018Fourier, Shenghui_method, grossman2026unbinnedextractiongammabto, Backus:2023NoBinning}, which aim to improve on the statistical precision of the binned method. The new methods are more complex than the binned method. To date, only one of these has been successfully applied to data~\cite{unbinned_GGSZ}, where the gain was found to be modest at 5$\%$, in part due to the complications arising in real data compared to the idealised expectations. A significant advantage of the binned method lies in its simplicity, and the power of the method is further augmented by the new schemes presented here. For \DtoKspipi decays, the value of $Q_\gamma=0.942$ in the \newgamma scheme suggests that further gains in unbinned variants are limited to $\sim6\%$. Similarly for \DtoKsKK decays, the \optkskkIII scheme is expected to retain $\sim 79\%$ of the statistical sensitivity in the unbinned limit, where the $\mathcal{N}=3$ variant closes the gap between the binned and potential unbinned analyses. With these gains, and the simplicity in performing the analysis, the binned method is expected to remain competitive. 

\section{Optimal binning for measurement of CP violation and mixing in charm}
\label{sec:charm_binning}
\subsection{Figure of merit for sensitivity to mixing and CP violation in charm}

The \DtoKspipi decay is also an important decay channel for measuring the charm mixing parameters and searching for time-dependent $CP$-violation in the $D$ meson system. In these analyses, $D^0$ and $\overline{D^0}$ mesons are separated into distinct samples by tagging their flavour at production. Assuming $CP$ symmetry, the time evolution of the decay distributions across the Dalitz plot is governed by the mixing and $D$ hadronic parameters. However, if $CP$ violation is present, either in mixing or in the interference between mixing and decay amplitudes, differences between the $D^0$ and $\overline{D^0}$ Dalitz plots can arise as a function of the $D^{0}$ decay time. Experimentally, mixing is most conveniently parameterised by $x_{CP}$ and $y_{CP}$, with $CP$-violating effects introduced by $\Delta x$ and $\Delta y$~\cite{Binflip}.

The most recent experimental measurements were performed using approaches similar to those used to measure $\gamma$, utilising the same external inputs to account for the dependence on \ci and \si. Within this framework, two distinct model-independent approaches exist. The first derives its sensitivity to the mixing parameters by fitting to the decay-time distribution in each individual Dalitz-plot bin~\cite{Bondar:2010qs,Thomas:2012qf}. The second approach, known as the ``bin-flip'' method, further bins the data into decay-time intervals and constructs time-dependent yield ratios between conjugate Dalitz bins ($i$ and $-i$)~\cite{Binflip}. Forming the ratio between kinematically similar regions of the Dalitz plot partially cancels both the decay-time and Dalitz-plot efficiencies, and also results in an improved sensitivity to $x_{CP}$ at the expense of $y_{CP}$. This method has been used by LHCb in its most precise measurements~\cite{Run2_binflip_prompt,Run2_binflip_sl}. To date, the bin-flip analysis has made use of the equal-phase binning scheme with 8 bins (\equalVIII) measured in Ref.~\cite{old_optimal_binning}. In the following, the concepts developed for optimising binning schemes for $\gamma$ are applied to the bin-flip analysis to extract greater sensitivity from the data.  

To define an analogous $Q$-metric for the charm case, first consider a likelihood function of the form,
\begin{equation}
    \ln \mathcal{L}(\theta) \approx -\frac{1}{2} \sum_{i} \frac{(R_i - \mu_i(\theta))^2}{\sigma_i^2} = -\frac{1}{2} \chi^2(\theta),
    \label{eq:lnL_binflip}
\end{equation}
as used in the bin-flip method~\cite{Binflip}, where $\theta$ is the physics parameter of interest and the observables $R_i$ are assumed to be Gaussian distributed. The bin-flip method is particularly sensitive to $x_{CP}$ and $\Delta x$. This, along with the fact that there are multiple other $D^{0}$ decay modes better suited for measuring $y_{CP}$ and $\Delta y$, makes $x_{CP}$ and $\Delta x$ the most interesting observables for this analysis. The decision is therefore made to optimise for $x_{CP}$, which, to a good approximation, is equivalent to optimising for $\Delta x$. 

By differentiating Eq.~\ref{eq:lnL_binflip} with respect to the observable of interest ($x_{CP}$) and taking the expectation value, the Fisher information follows as,
\begin{equation}
    \mathcal{I}(x_{CP}) = -\mathbb{E}\!\bigg[\frac{\partial^2}{\partial x_{CP}^2}\ln\mathcal{L}\bigg] = \sum_{i} \frac{1}{\sigma_i^2} \left( \frac{\,\partial \mu_i(x_{CP})}{\,\partial x_{CP}} \right)^2,
\end{equation}
which is analogous to Eq.~\ref{eq:Fisher}, but with $N_i$ replaced by $\sigma_i^2$. For simplicity, $CP$ conservation is assumed, and a small mixing approximation is applied such that only terms linear in the mixing parameters ($x_{CP}, y_{CP} \sim 0.005$) are retained. Integrating over time and simplifying produces the final $Q$-metric
\begin{equation}
    Q_{x_{CP}}^2 = \frac{\displaystyle\sum_i \Big(K_{+i} + K_{-i}\Big)\, s_i^2}
    {\displaystyle\int\Big(|A_{D}(m_{+}^{2}, m_{-}^{2})|^2 + |A_{D}(m_{-}^{2}, m_{+}^{2})|^2\Big)\sin^{2}\Delta\delta_{D}(m_{+}^2, m_{-}^2)\,\dd m_{+}^{2}\,\dd m_{-}^{2}\,},
    \label{eq:Q_binflip}
\end{equation}
where the numerator is $\mathcal{I}(x_{CP})$ for the binned case. The denominator, $\mathcal{I}(x_{CP})|_{\mathcal{N}\to\infty}$, follows analogously for the unbinned limit by defining the ratio to be between a specific point on the Dalitz plot and its mirror point across the diagonal.

\begin{figure}[ht!]
    \centering
    \begin{subfigure}[b]{0.49\textwidth}
        \centering
        \includegraphics[width=\textwidth]{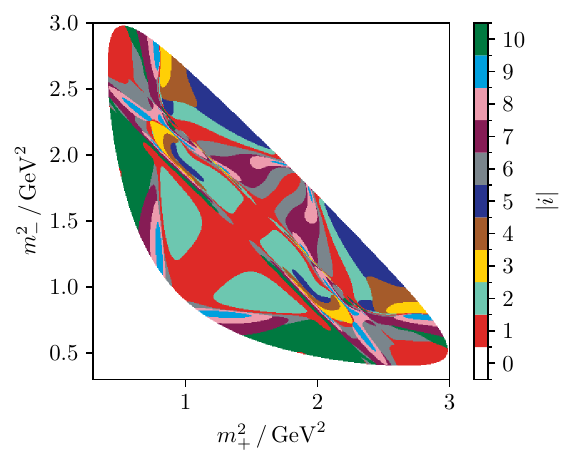}
        \caption{Optimised using Eq. \ref{eq:Q_binflip}, $Q_{x_{CP}}=0.957$}
    \end{subfigure}
    \hfill
    \begin{subfigure}[b]{0.49\textwidth}
        \centering
        \includegraphics[width=\textwidth]{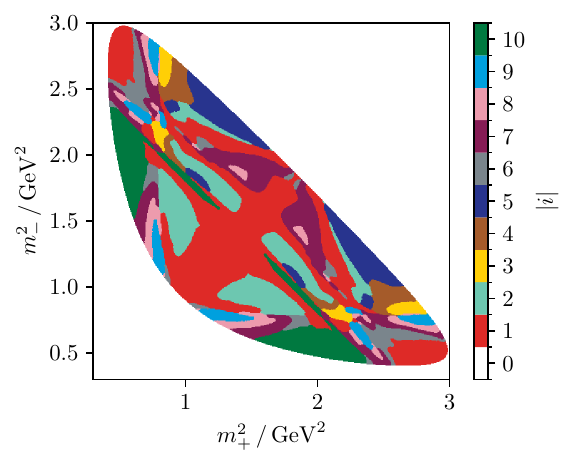}
        \caption{Optimised for Eq. \ref{eq:Q_binflip_with_migration}, $Q_{x_{CP}}=0.925$}
    \end{subfigure}
    \caption{The $\mathcal{N}=10$ binning scheme optimised for Eq. \ref{eq:Q_binflip} along with the \newcharm scheme optimised for Eq. \ref{eq:Q_binflip_with_migration}.}
    \label{fig:charm_binning}
\end{figure}

The binning scheme with $\mathcal{N}=10$ is depicted in Fig.~\ref{fig:charm_binning}. The increase in $Q$-metric with respect to the \equalVIII binning scheme indicates an approximate $20\%$ improvement in statistical sensitivity. The size of the systematic uncertainty arising from bin migration is also important as it poses a greater concern in the bin-flip analysis than in $\gamma$ measurements, where the effect is accounted for to first order by using the high-yielding $B^{\pm}\rightarrow D\pi^{\pm}$ decay as a normalisation mode~\cite{Run12_BPGGSZ, paper_Innes, paper_Seophine}. Although a smoothing procedure is introduced, its effectiveness is limited, as will be seen in the following sections. Because the sub-bin size is comparable to the $m_{\pm}$ resolution, removing small artefacts has a negligible impact beyond visual aesthetics. Ultimately, bin migration is highly sensitive to the placement of bin boundaries relative to the underlying Dalitz-plot intensity. Therefore, additional measures must be taken to minimise this effect.

No other experimental effects are included in the optimisation. As efficiency variations across the Dalitz plot were found to have a negligible effect during the $\gamma$ binning optimisation (Sec.~\ref{sec:detector_efficiencies}), the same approach is adopted here. Furthermore, decay-time acceptance and finite decay-time resolution are not expected to significantly impact the choice of an optimal Dalitz binning. Finally, the high signal purity of these data samples makes it unnecessary to consider background components.

\subsection{Pseudoexperiments}

Pseudoexperiments are used to assess the statistical uncertainty and the size of the bias arising from bin migration effects. The pseudodatasets are generated in a similar manner to those for $\gamma$. The binned yields, $N_{ij}^{\pm}$, for each Dalitz bin $i$ and decay-time bin $j$ are predicted from the time-dependent rate equations~\cite{Binflip}. The expected values for $K_i$, $c_i$, and $s_i$ are derived from the amplitude model~\cite{Belle2018}, while the mixing parameters are taken from the latest LHCb combination~\cite{2024_combination}, and $CP$ conservation is assumed. To incorporate the effect of bin migration, the expected yields are modified using the predicted net migration. Aside from migration, no other experimental effects are included in the generation process.

For each pseudodataset, the yields are sampled from a Poisson distribution with expectation values $N_{ij}^{\pm}$. The pseudodata are generated with an average yield of 200 million events, corresponding to the estimated size of an LHCb Run 3 analysis~\cite{LHCb:2018roe}. A total of 5000 pseudodatasets are generated and processed following the standard bin-flip approach~\cite{Binflip}. The \ci and \si parameters are fixed in the fit. However, the $r_i\equiv K_{-i}/K_{i}$ ratios are left as free parameters, since the effect of migration is large enough that fixing them to their expected values would otherwise bias the mixing measurement. This is consistent with LHCb analyses, where these ratios are typically left free in the fit~\cite{Run2_binflip_prompt}.

\subsection{Minimising bias due to bin migration}
\label{sec:bin_migration}
\begin{figure}
    \centering
    \includegraphics[width=\linewidth]{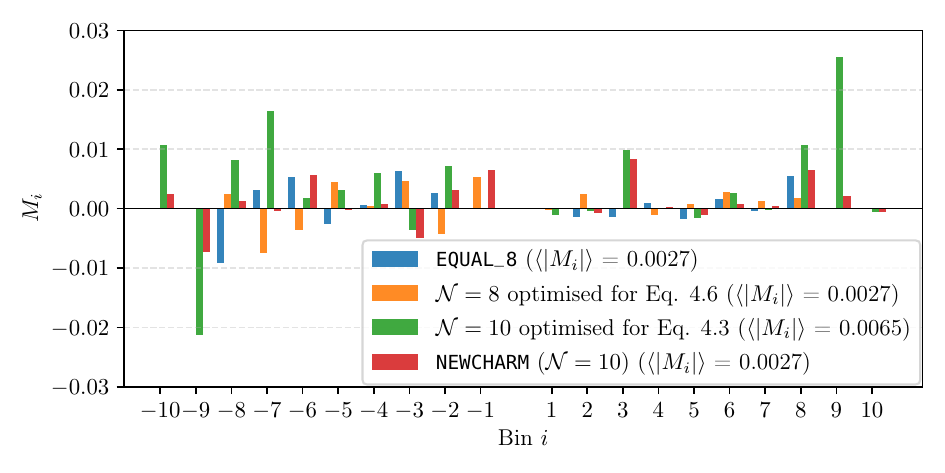}
    \caption{Estimated net migrations for different binning schemes considered in this study, alongside the \equalVIII scheme currently in use.}
    \label{fig:net_migration_charm}
\end{figure}
The $M_i$ are estimated in the same way as described in Section~\ref{sec:systematics} and are shown in Fig.~\ref{fig:net_migration_charm}, from which it is seen that the binning scheme optimised using Eq. \ref{eq:Q_binflip} exhibits significantly larger net migration, $M_i$, than the \equalVIII binning scheme. The resultant bias in the relevant observables as determined by pseudoexperiments is shown in Table~\ref{tab:bin_migration_systematic}. This bias is almost twice the size of that observed for the \equalVIII scheme, and is equivalent to $\sim 70\%$ of the statistical uncertainty. This is problematically large and would preclude the use of this binning scheme in favour of existing schemes. 

A new metric is therefore defined, which includes a penalty factor designed to reduce the amount of bin migration. In order to minimise computational expense, only the probability of migration from a sub-bin to its nearest neighbour is considered in the metric computation. The net migration $M_i$ can be computed and a global quantity $M$ is taken as proxy for the degree of migration,
\begin{equation}
    M=\displaystyle\sum_i \big|M_i\big|.
\end{equation}
The penalty term is then defined as
\begin{equation}
    Q_M^2=\frac{M_0^2}{M^2},
    \label{eq:Q_migration}
\end{equation}
where $M_0$ is the total migration, $M$, of the \equalVIII scheme. The penalty factor is then combined with Eq.~\ref{eq:Q_binflip} to define a metric that takes into account both statistical precision and the level of migration,
\begin{equation}
    Q_{x_{CP},M}^2=\alpha Q_{x_{CP}}^2+(1-\alpha)Q_M^2.
    \label{eq:Q_binflip_with_migration}
\end{equation}

The starting point of the optimisation is the scheme derived using Eq.~\ref{eq:Q_binflip}. A penalty weight of $\alpha=0.997$ is chosen empirically to maximise the statistical sensitivity, $Q_{x_{CP}}$, while reducing the global migration, $M$, to a level comparable with that of the \equalVIII scheme. Due to the nearest-neighbour approximation, optimising using Eq.~\ref{eq:Q_binflip_with_migration} is found to result in small structures and artefacts. These are minimised by setting $f_{\mathrm{grow}}$ to zero when optimising with Eq.~\ref{eq:Q_binflip_with_migration}. Furthermore, because this scheme produces many irregular bin boundaries, a harsher smoothing procedure is applied than the nominal. An initial round of the algorithm is run with $r=3$ and $n_{\mathrm{cut}}=0.75$ and then remaining structures smaller than 5 sub-bins are removed. 

The optimisation process is performed for both $\mathcal{N}=8$ and $10$. The net migrations, $M_i$, for the resulting binning schemes are shown in Fig.~\ref{fig:net_migration_charm}. The expected statistical and systematic uncertainties for various schemes, highlighting the impact of these different metric choices, are presented in Table~\ref{tab:bin_migration_systematic}. The improvement in statistical uncertainty between the $\mathcal{N}=8$ and $\mathcal{N}=10$ equal-phase binning schemes is evident. However, the dedicated charm optimisation yields significantly larger improvements of approximately 20$\%$. Interestingly, while the \newgamma scheme also provides a distinct improvement in statistical uncertainty, it performs poorly with respect to the estimated systematic uncertainty due to migration. Conversely, including the migration penalty term in the charm optimisation using Eq.~\ref{eq:Q_binflip_with_migration} effectively reduces the associated systematic bias while preserving statistical sensitivity. These migration-aware schemes are therefore considered further.

\begin{table}[htbp]
    \centering
     \caption{Summary of statistical uncertainties and biases due to bin migration in units of $10^{-3}$, obtained with the different binning schemes. The statistical uncertainties are obtained for a sample of 200M signal decays. The relative uncertainties on the reported values of $\sigma_{\textrm{stat}}$ and $\sigma_{\textrm{migration}}$ are $\sim 1\%$.}
    
    \footnotesize
    \begin{tabular}{lcccc}
    \\
        \toprule
        Binning Scheme & $\sigma_{\textrm{stat}}/\sigma_{\textrm{\equalVIII}} $ & $\sigma_{\textrm{stat}}$  &  $\sigma_{\textrm{migration}}$  & $\sigma_{\textrm{migration}}/\sigma_{\textrm{stat}}$ \\
        \midrule
        \equalVIII &  1.000  &  0.156  &   0.035  &  0.226 \\
        $\mathcal{N}=8$ optimised with Eq. \ref{eq:Q_binflip} &  0.777  &  0.121  &   0.088  &  0.726 \\
        $\mathcal{N}=8$ optimised with Eq. \ref{eq:Q_binflip_with_migration} &  0.815  &  0.127  &   0.039  &  0.308  \\
        \midrule
        \equalX &  0.953  &  0.149  &   0.042  &  0.286 \\
        $\mathcal{N}=10$ optimised with Eq. \ref{eq:Q_binflip} &  0.770  &  0.120  &   0.080  &  0.667 \\
        $\mathcal{N}=10$ optimised with Eq. \ref{eq:Q_binflip_with_migration}, \newcharm &  0.801  &  0.125  &   0.045  &  0.359 \\
         \newgamma &  0.851  &  0.133  &   0.070  &  0.531 \\
        
        \bottomrule
    \end{tabular}
    
    \label{tab:bin_migration_systematic}
\end{table}

It is interesting to note that moving from $\mathcal{N}=8$ to $10$ bins yields only a marginal sensitivity gain, making the argument for increasing $\mathcal{N}$ less immediately clear. Considering systematic uncertainties, the estimated bias from bin migration is smaller for the $\mathcal{N}=8$ scheme. On the other hand, some systematic uncertainties are expected to improve with increased $\mathcal{N}$. As demonstrated in Sec.~\ref{sec:strong_phase_systematic}, the uncertainty on $\gamma$ from the strong phases generally decreases with more bins. Because this is expected to be a non-negligible source of uncertainty in the full Run 1--3 measurements, the $\mathcal{N}=10$ scheme is proposed as the optimal choice for the bin-flip analysis, and is denoted as \newcharm. The scheme is shown in Fig.~\ref{fig:charm_binning}.

The pseudoexperiments show approximately the same gain in sensitivity for $\Delta x$, confirming that optimising for $x_{CP}$ is, to a good approximation, equivalent. It is also worth considering the effect on the sensitivity to $y_{CP}$, even though measuring it is not the primary goal of the bin-flip method. Relative to the \equalVIII binning scheme, the uncertainty on $y_{CP}$ increases by $\sim35\%$ for the \newcharm scheme and $\sim115\%$ for the analogous $\mathcal{N}=8$ scheme. This better preservation of sensitivity to $y_{CP}$ serves as a further argument for selecting the $\mathcal{N}=10$ scheme. The \newcharm scheme is also tested for its applicability to $\gamma$ measurements, but unfortunately is found to increase the statistical uncertainty by more than 30$\%$ compared to the \cleooptimal scheme. 

\section{Conclusions}
This paper defines new binning schemes for \DtoKspipi and \DtoKsKK decays. The exact specification of these can be found in Ref.~\cite{binning_schemes}. The scheme optimised for sensitivity to $\gamma$ in $B^{\pm}\rightarrow DK^{\pm}$ decays with \DtoKspipi provides a gain of approximately 5$\%$ in the expected sensitivity to $\gamma$ in $B^\pm \to D K^\pm$ decays relative to the \cleooptimal scheme used in recent analyses~\cite{Run12_BPGGSZ, Run3_BPGGSZ, paper_Seophine, BESIII_8fb-1}. This gain is primarily provided by a new figure of merit that better represents the sensitivity to $\gamma$, alongside a more accurate treatment of the relevant background levels and composition. Part of this gain also comes from an increase in the number of bins to $\mathcal{N}=10$, which is chosen balancing the gains in statistical precision with considerations of applicability to measurements of the strong-phase differences at BESIII. In addition, relevant systematic uncertainties are studied. The propagated uncertainty in $\gamma$ due to the strong-phase inputs is expected to reduce by 10$\%$. The bin migration due to detector resolution in the \newgamma scheme is also shown to be similar to that of \cleooptimal. The use of the \newgamma scheme is not expected to be detrimental in other $B$ decay channels in comparison to continued use of the \cleooptimal scheme.  

This formalism is extended to the \DtoKsKK decay, for which optimal binning schemes with $\mathcal{N}=2, 3$, and 4 are presented, with the expectation that only one of these is adopted for use. An additional augmented figure of merit is used, which takes into account correlation between $\gamma$ and the strong-phase parameter, $\delta_B$, which is prevalent in schemes with few bins. The gain in expected sensitivity by using the \optkskkIII scheme for \DtoKsKK decays instead of the $\mathcal{N}=2$ equal-phase schemes is estimated in pseudoexperiments to be approximately 12$\%$.

In addition, for the first time, a binning scheme optimised for sensitivity to observables related to mixing and $CP$ violation in the charm system is presented. Because the bin-flip analysis is more sensitive to bin migration and prone to larger resultant biases, a penalty term is introduced into the figure of merit (Eq.~\ref{eq:Q_binflip_with_migration}) to minimise this effect. This results in a bias comparable to that of the \equalVIII scheme previously in use~\cite{Run2_binflip_sl, Run2_binflip_prompt}, while retaining most of the potential improvements in statistical precision. This scheme is expected to yield a statistical precision on $\Delta x$/$x_{CP}$ that is 20$\%$ better than that of the \equalVIII scheme previously in use~\cite{old_optimal_binning, Run2_binflip_prompt, Run2_binflip_sl}.

For \DtoKspipi decays, two schemes have been proposed, one that targets measurements of the CKM angle $\gamma$ and the other that targets measurements of mixing and indirect $CP$ violation in neutral charm decays. Recent combinations have brought these two bodies of work together~\cite{simgamcharm,2024_combination} and the use of two different binning schemes would prevent the full combination in the \DtoKspipi channel, resulting in the double counting of the strong-phase related uncertainties. However, the current immediate goals in both streams of analysis are statistical precision and the observation of indirect $CP$ violation in neutral charm decays; the use of two separate binning schemes maximises this impact. In the far future, when systematic uncertainty correlations between these two analyses become important, a change in the bin-flip analyses can be envisaged where migration corrections form a core part of the analysis, thus reducing the problematic systematic uncertainty. Under such a regime, the \newgamma scheme could also be used with good sensitivity to the charm physics analyses.

\acknowledgments

The authors are grateful for discussion and feedback from G. Wilkinson, S. Zhang and S. Stanislaus. M. Bovill and S. Malde thank STFC for financial support. M. Bovill also thanks the Clarendon Fund and the Jowett Scholarship of Balliol College.

%\paragraph{Note added.} This is also a good position for notes added
%after the paper has been written.

% Bibliography

%% [A] Recommended: using JHEP.bst file
%% \bibliographystyle{JHEP}
%% \bibliography{biblio.bib}

%% or
%% [B] Manual formatting (see below)
%% (i) We suggest to always provide author, title and journal data or doi:
%% in short all the informations that clearly identify a document.
%% (ii) please avoid comments such as "For a review'', "For some examples",
%% "and references therein" or move them in the text. In general, please leave only references in the bibliography and move all
%% accessory text in footnotes.
%% (iii) Also, please have only one work for each \bibitem.

% Bibliography using BibTeX (remove the following if you prefer manual \bibitem)
\bibliographystyle{JHEP} % or another bst available to you
\bibliography{biblio}    % expects biblio.bib in working directory

\end{document}